\documentclass[twocolumn]{jpsj3}
\def\runauthor{D. Aoki and J. Flouquet}

\title{Ferromagnetism and Superconductivity in Uranium Compounds}

\author{%
Dai~\textsc{Aoki}$^1$\thanks{E-mail address: dai.aoki@cea.fr} %
and Jacques~\textsc{Flouquet}$^1$\thanks{J.F. is also ``directeur de recherche \'{e}m\'{e}rite'' in CNRS}
}

\inst{%
$^1$INAC/SPSMS, CEA-Grenoble, 17 rue des Martyrs, 38054 Grenoble, France\\
}

\abst{%
Recent advances on ferromagnetic superconductors, UGe$_2$, URhGe and UCoGe are presented.
The superconductivity (SC) peacefully coexists with the ferromagnetism (FM), 
forming the spin-triplet state of Cooper pairs.
The striking new phenomena, such as SC reinforced by the magnetic field, are
associated with Ising-type ferromagnetic fluctuations.
A variety of ferromagnetic ordered moments between UGe$_2$, URhGe and UCoGe
affords to understand the relation between FM, tricriticality and SC.
}

\kword{ferromagnetism, superconductivity, spin-triplet state, magnetic fluctuation, UGe$_2$, URhGe, UCoGe}

\begin{document}
\maketitle

\section{Introduction}

The ferromagnetism had been thought to be antagonistic to superconductivity (SC) 
in the framework of singlet $s$-wave pairing up to the discovery of new class of materials like
the Chevrel phase compounds (REMo$_6$Se$_8$, RE: rare earth).~\cite{Fis90_FM}
Antiferromagnetism (AF) obviously coexists peacefully with SC in these RE intermetallic compounds 
where the $4f$ electrons are localized on the RE site
and the magnetic interaction between RE ions are mediated indirectly by the conduction electrons 
via so-called RKKY (Ruderman, Kittel, Kasuya, Yosida) interaction.
The simple image is that under the large superconducting coherence length $\xi_0$,
the average value of the magnetization is zero 
as the AF periodicity is generally smaller than $\xi_0$.~\cite{Flo02}
In contrast, no microscopic coexistence of FM and SC has been observed.
In HoMo$_6$Se$_8$~\cite{Ish77} or ErRh$_4$B$_4$~\cite{Fer77}, SC is observed in the intermediate temperature range approximately from $2$ to $10\,{\rm K}$,
but when it is cooled further, FM destroys SC at low temperatures.
For example in ErRh$_4$B$_4$, SC appears below $8.7\,{\rm K}$.
Because of the internal conflicts between SC and FM, 
an intermediate phase exists in the narrow temperature range between $0.8\,{\rm K}$ and $1\,{\rm K}$
with the establishment of a modulated magnetic structure which can be regarded as a domain-like arrangement with 
the periodicity $d \ll \xi_0$.
However, further cooling, FM is overwhelmed and $s$-wave SC is destroyed as basically the energy gained by the RE atoms due to
the magnetic transition ($k_{\rm B}T_{\rm Curie}$) is far higher than the energy gained by the electrons as they form Cooper pairs 
at the SC transition ($(k_{\rm B}T_{\rm sc})^2 \rho (\varepsilon_{\rm F})$),
where $\rho (\varepsilon_{\rm F})$ is the density of states of the conduction electrons at the Fermi level.
In the case of AF--SC systems, the unusual situation can happen under magnetic field,
when the applied magnetic field may cancel the total internal field and thus
the field re-entrant SC is observed as predicted by the so-called Jaccario-Peter compensation effect.~\cite{Jac62}
This was first observed in the Chevrel phase compound~\cite{Meu84} and recently in the organic systems $\lambda$(BETS)$_2$FeCl$_4$.~\cite{Bal01,Uji01}

\subsection{Unconventional superconductivity}
In the previous case, the electrons which are responsible for the magnetism are basically localized.
In strongly correlated electron systems such as heavy fermion compounds, 
the high $T_{\rm c}$ cuprates and the new Fe-pnictide families, at first approximation,
the electrons must be regarded as itinerant and thus the interplay between magnetism and SC will be strong.~\cite{Flo06_review}
Furthermore, the importance of the Coulomb repulsion between the electrons wipes out 
the possibility of $s$-wave pairing based on the BCS theory 
because the attractive force mediated by phonon cannot overcome the strong Coulomb repulsive force.
This situation opens the opportunity to search for unconventional superconductivity 
with $d$-wave or $p$-wave pairings possessing finite angular momentum
leading to novel SC properties such as anisotropic SC gap and order parameters with
specific temperature and magnetic field response.
Here, an important reference is the triplet $p$-wave superfluidity of the Fermi liquid $^3$He~\cite{Leg75}
with exotic $A$ phase and $B$ phase;
the former one characterized by the so-called equal spin pairing with separation between $\uparrow\uparrow$
and $\downarrow\downarrow$ spin carriers.
Magnetic field can even give rise to two successive superfluid transition.
In contrast to the case of the uranium ferromagnetic superconductors,
the liquid phase of $^3$He never reach a ferromagnetic instability as shown in the weak pressure dependence of
i) its Landau parameter $F_0^a$ describing the enhancement of the susceptibility by comparison to the specific heat and
ii) its weak Gr\"{u}neisen parameter~\cite{Flo82}

The great advantage of the heavy fermion compounds based on the presence of 4$f$ (cerium, ytterbium) or 5$f$ (uranium, neptunium, plutonium)
electrons, which are quite ready to become magnetic, is that
moderate pressure and even magnetic field can drive them to magnetic instability i.e. 
from long range magnetic order (AF or FM) to paramagnetic (PM) ground state at a critical pressure $P_{\rm c}$.
As now the Cooper pair mechanism is linked to the electron correlation itself,
SC will be often observed in a dome shape centered around $P_{\rm c}$,
where the magnetic fluctuation is enhanced.

Starting with the discovery of CeCu$_2$Si$_2$ superconductivity in 1979~\cite{Ste79}
there are now many macroscopic as well as microscopic evidences of $d$-wave spin singlet superconductivity 
in the heavy fermion compounds interplaying between SC and AF.

The coexistence of FM and SC was first discovered in UGe$_2$\cite{Sax00} under pressure in 2000, almost two decades after the discovery of SC in CeCu$_2$Si$_2$.
Soon afterward, the SC was found in the weak ferromagnet URhGe for the first time at ambient pressure.~\cite{Aok01} 
Recently UCoGe with identical crystal structure of URhGe was found to be a ferromagnetic superconductor, as well.~\cite{Huy07}
In all of these compounds, $T_{\rm sc}$ is lower than $T_{\rm Curie}$,
indicating that SC phase exists in the FM phase, which is contrary to the
previous case such as ErRh$_4$B$_4$ or Chevrel phase compounds where $T_{\rm sc}$ is higher than $T_{\rm Curie}$.
Furthermore, the ordered moments of uranium ferromagnetic superconductors are 
much lower than those expected from the free uranium ion.
Therefore 5$f$ electrons are naively believed to be itinerant.
To date, all the ferromagnetic superconductors are uranium compounds. 
The well-known weak ferromagnet ZrZn$_2$ was first reported to reveal SC,~\cite{Pfl01} however 
after careful sample preparations and characterizations, SC was found to be extrinsic most likely due to the
Zr alloys on the surface.~\cite{Yel05}
SC is observed in the ferromagnet UIr with non-inversion symmetry of the crystal structure in FM3 phase
in narrow pressure range ($2.6 \lesssim P \lesssim 2.8\,{\rm GPa}$) with the maximum $T_{\rm sc}\sim 0.15\,{\rm K}$.
The bulk SC has not been established yet. 
The upper critical field of SC is quite small ($\sim 0.026\,{\rm T}$)
compared to those for above mentioned three ferromagnetic superconductors.

In this paper, first we review experimental results of three ferromagnetic superconductors UGe$_2$, URhGe and UCoGe.
Next we describe some theoretical views for FM and SC. 
Finally the conclusion and remarks are given.
A very recent complementary our review of ferromagnetic superconductors can be found in Ref.~\citen{Aok11_CR}

\section{UGe$_2$--the first ferromagnetic superconductor: superconductivity and phase diagram}
\begin{figure}[tbh]
\begin{center}
\includegraphics[width=1 \hsize,clip]{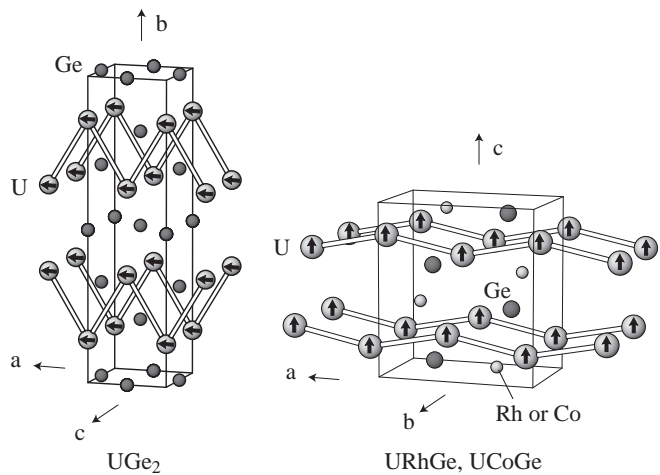}
\end{center}
\caption{Crystal structure of UGe$_2$, URhGe and UCoGe. The arrows on the U site denote the direction of the moment.}
\label{fig:Structure}
\end{figure}

\begin{figure}[tbh]
\begin{center}
\includegraphics[width=1 \hsize,clip]{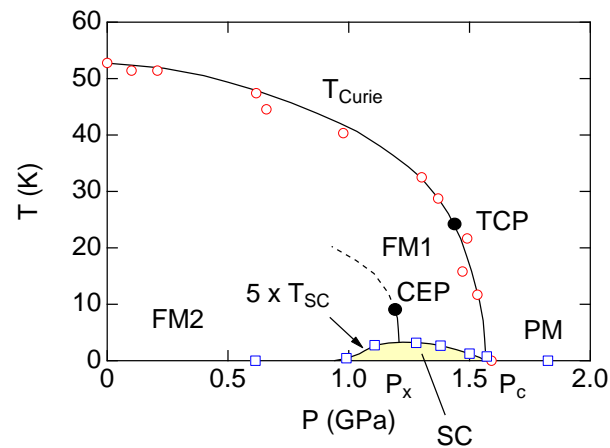}
\end{center}
\caption{(Color online) Temperature--pressure phase diagram of UGe$_2$. FM1, FM2 and PM represent the ferromagnetic state with large moment ($\sim 1.5\,mu_{\rm B}$), the ferromagnetic state with small moment ($\sim 1\,\mu_{\rm B}$) paramagnetic state, respectively. Above the pressure of the tricritical point (TCP), the first order ferromagnetic transition is observed. Below the pressure of critical end point for $T_x$, the crossover occurs between FM1 and FM2.~\protect\cite{Sax00,Tau11,Hux01}}
\label{fig:UGe2_Saxena}
\end{figure}

UGe$_2$ crystallizes in the orthorhombic structure, as shown in Fig.~\ref{fig:Structure}.~\cite{Oik96}
The U zig-zag chain with the distance of the next nearest neighbor $d_{\mbox{U-U}} = 3.85\,{\rm \AA}$ is formed along $a$-axis, which is similar to $\alpha$-U with CDW.
The FM transition had been observed at $T_{\rm Curie}=52\,{\rm K}$~\cite{Onu92}
and the ordered moment is relatively large, $M_0 \sim 1.5\,\mu_{\rm B}$.
The properties of UGe$_2$, together with URhGe and UCoGe are summarized in Table~\ref{tab:table1}.
The magnetic moment is directed along $a$-axis.
With increasing pressure $T_{\rm Curie}$ collapses and finally PM ground state is realized above the critical pressure $P_{\rm c}\sim 1.5\,{\rm GPa}$~\cite{Oom95}.
Surprisingly, SC appears around $1.2\,{\rm GPa}$ with $T_{\rm sc}\sim 0.7\,{\rm K}$ as a maximum.
As shown in Fig.~\ref{fig:UGe2_Saxena}, at this pressure $T_{\rm Curie}\sim 35\,{\rm K}$ is much higher than $T_{\rm sc}$ and 
$M_0$ is also large ($\sim 1\,\mu_{\rm B}$).
\begin{table}[tbhp]
\begin{center}
\caption{Characteristic properties of UGe$_2$, URhGe and UCoGe. 
(%
$d_{\mbox{U--U}}$: distance of the first nearest neighbor of U atom,
$M_0$: ordered moment,
$H_{\rm int}$: internal field associated with $M_0$,
$P_{\rm c}$: critical pressure between FM and PM,
$H_{\rm c2}^{a,b,c}$: upper critical field for $H\parallel a,b,c$-axis)
$^\dagger$the values of $H_{\rm c2}$ in UGe$_2$ are at $\sim 1.2\,{\rm GPa}$.
}
\begin{tabular}{lccc}
\hline
								& UGe$_2$		& URhGe		& UCoGe								\\
\hline
Structure						& Ortho.			& Ortho.		& Ortho.								\\
Space group						& $Cmmm$			& $Pnma$		& $Pnma$								\\
$d_{\mbox{U--U}}\,({\rm \AA})$	& 3.85			& 3.50		& 3.48								\\
$T_{\rm Curie}\,({\rm K})$		& 52				& 9.5		& $\sim 3$							\\
$M_0\,(\mu_{\rm B})$				& 1.48			& 0.4		& $\sim 0.05$						\\
Mag. easy-axis					& $a$			& $c$		& $c$								\\
$H_{\rm int}\,({\rm T})$			& 0.28			& 0.08		& 0.01								\\	
$\gamma\,({\rm mJ/K^2 mol})$		& 34				& 160		& 55									\\
$P_{\rm c}\,({\rm GPa})$			& 1.5			& $<0$		& $\sim 1.2$							\\
$T_{\rm sc}\,({\rm K})$			& 0.8			& 0.26		& 0.7								\\
$\Delta C/\gamma T_{\rm sc}$		& $\sim 0.3$		& 0.6		& 0.7								\\
$H_{\rm c2}^a\,({\rm T})$		& 1.4$^\dagger$	& 2.5		& $>30$								\\
$H_{\rm c2}^b\,({\rm T})$		& 2.4$^\dagger$	& 2			& $\sim 18$							\\
$H_{\rm c2}^c\,({\rm T})$		& 4.8$^\dagger$	& 0.7		& 0.6								\\
\hline
\end{tabular} 
\end{center}
\label{tab:table1}
\end{table}

\begin{figure}[tbh]
\begin{center}
\includegraphics[width=0.7 \hsize,clip]{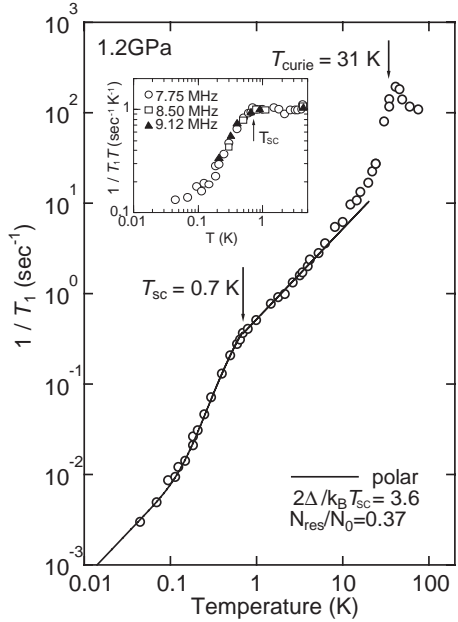}
\end{center}
\caption{Temperature dependence of the spin-lattice relaxation rate $1/T_1$ by NQR experiments at $1.2\,{\rm GPa}$ in UGe$_2$, as an evidence of microscopic coexistence of FM and SC.~\protect\cite{Kot05}}
\label{fig:UGe2_T1}
\end{figure}

\begin{figure}[tbh]
\begin{center}
\includegraphics[width=0.8 \hsize,clip]{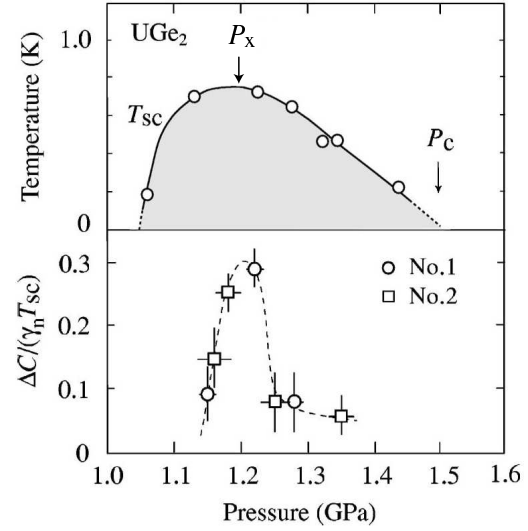}
\end{center}
\caption{Pressure variation of $T_{\rm sc}$ by resistivity measurements and the $\Delta C/\gamma T_{\rm sc}$ by specific heat measurements in UGe$_2$.~\protect\cite{Tat04}}
\label{fig:UGe2_SC_phase}
\end{figure}

Both $T_{\rm Curie}$ and $M_0$ collapse at $P_{\rm c}$.
Complementary measurements~\cite{Hux00,Hux01} show later that the maxima of $T_{\rm sc}$ corresponds to the pressure just at $P_{\rm x}\sim 1.2\,{\rm GPa}$
where the system switches from 
large moment ($M_0\sim 1.5\,\mu_{\rm B}$) at low pressure phase (FM2) to
small moment ($M_0\sim 1  \,\mu_{\rm B}$) at high pressure phase (FM1), through a first order transition.
The transition from FM1 to PM is also associated with the first order transition at $P_{\rm c}$ with
an abrupt drop of sublattice magnetization ($\Delta M_0\sim 0.8\,\mu_{\rm B}$).~\cite{Pfl02}
Evidences for homogeneous coexistence of FM and SC at $P\sim P_{\rm x}$ were given by
the persistence of FM in the SC phase observed in neutron diffraction experiments~\cite{Hux03},
the temperature dependence of nuclear spin-lattice relaxation rate in NQR measurements (see Fig.~\ref{fig:UGe2_T1})~\cite{Kot05}
and the specific heat jump at $T_{\rm sc}$.~\cite{Tat01}
Open question is how the homogeneous coexistence of FM and SC is realized below and above $P_{\rm x}$.
Figure~\ref{fig:UGe2_SC_phase} shows the pressure variation of $T_{\rm sc}$ obtained by the resistivity measurements
and the $\Delta C/\gamma T_{\rm sc}$ by the specific heat measurements.~\cite{Tat01,Tat04}
$T_{\rm sc}$ shows the maximum at $P_{\rm x}$.
The specific heat jump is much smaller than the value expected for weak coupling BCS scheme.
Furthermore the residual Sommerfeld coefficient ($\gamma$-value) is quite large, $70\,{\%}$ of $\gamma$-value in the normal state, in spite of very high quality single crystal.
This might be related with the large sublattice moment and the self-induced vortex state.
Another possible reason is the first order transition between FM1 and FM2, 
which can induce the phase separation of FM1 and FM2 due to the the small pressure gradient
in the pressure cell as well as in the the sample.

\begin{figure}[tbh]
\begin{center}
\includegraphics[width=1 \hsize,clip]{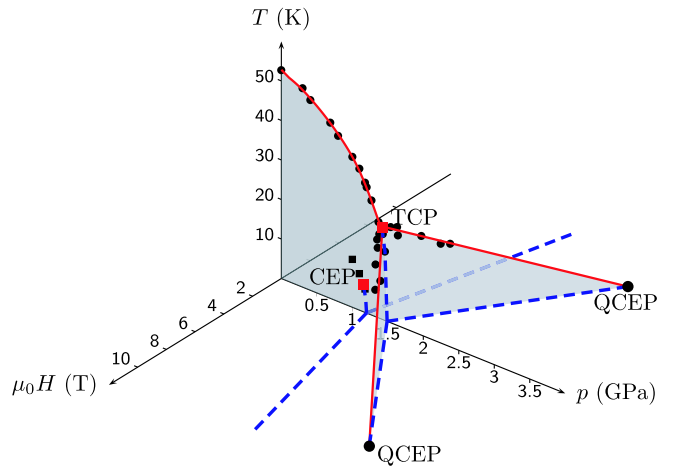}
\end{center}
\caption{(Color online) Temperature-pressure-field phase diagram of UGe$_2$ for $H\parallel M_0$ ($a$-axis).~\protect\cite{Tau10,Kot11}}
\label{fig:UGe2_TPH}
\end{figure}

Another tool to modify the $P_x$ and $P_{\rm c}$ boundary is to apply the magnetic field along the magnetization easy-axis ($a$-axis).
At the pressure of FM1 phase at zero field ($P_{\rm x}<P<P_{\rm c}$), 
the FM2 phase is recovered through the metamagnetic transition at $H=H_{\rm x}$.
In the PM phase, just above $P_{\rm c}$, a cascade of first order metamagnetic transition occurs at $H_{\rm c}$ (from PM to FM1) and 
$H_{\rm x}$ (from FM1 to FM2).
Careful studies for $H\parallel M_0$ ($a$-axis) was recently realized in order to clarify the FM-PM border of UGe$_2$
as it is an excellent example for tricriticality in itinerant ferromagnet.~\cite{Tau10,Kab10}
Under pressure, at $H=0$, the phase transition changes from second order to first order at
a tricritical point $T_{\rm TCP}=24\,{\rm K}$, $P_{\rm TCP}\sim 1.42\,{\rm GPa}$,
which is very close to $P_{\rm c}=1.49\,{\rm GPa}$.~\cite{Kab10,Tau11}
Under magnetic fields above $P_{\rm c}$, the first order metamagnetic transition will terminate at a critical end point
($P_{\rm QCEP}\sim 3.5\,{\rm GPa}$, $H_{\rm QCEP}\sim 16\,{\rm T}$, see Fig.~\ref{fig:UGe2_TPH})~\cite{Tau10,Kot11}
For the transition between FM1 and FM2, the critical end point at $H=0$ is located 
at $T_{\rm CEP}^{\rm x}\sim 7\,{\rm K}$ and $P_{\rm CEP}^{\rm x}\sim 1.16\,{\rm GPa}$,
which is very near the pressure where SC dome is suppressed.
Below $P_{\rm CEP}^{\rm x}$, $T_{\rm x}$ is a crossover between FM1 and FM2.

\begin{figure}[tbh]
\begin{center}
\includegraphics[width=1 \hsize,clip]{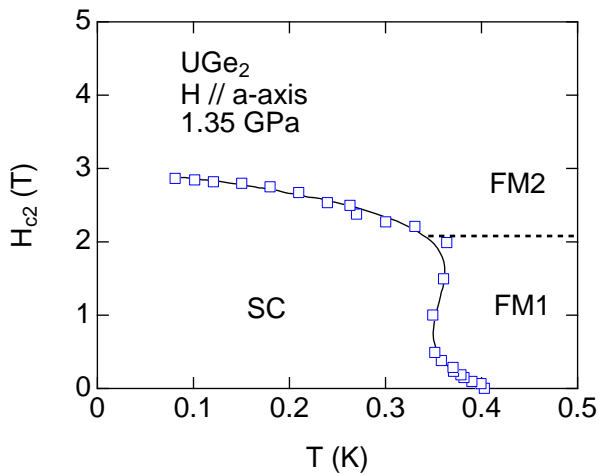}
\end{center}
\caption{(Color online) Temperature dependence of $H_{\rm c2}$ for $H\parallel a$-axis in UGe$_2$ at $1.35\,{\rm GPa}$, which is just above $P_{\rm x}$. The metamagnetic transition is detected at $H_x$ between FM1 and FM2.~\protect\cite{She01}}
\label{fig:UGe2_Hc2_a_axis}
\end{figure}

A striking point on the SC phase is that
the temperature dependence of the superconducting upper critical field $H_{\rm c2}$ for $H\parallel a$-axis (easy magnetization axis) 
indicates the field-enhanced SC phase 
when the pressure is tuned just above $P_{\rm x}$, as shown in Fig.~\ref{fig:UGe2_Hc2_a_axis}.~\cite{She01}
This peculiar shape of $H_{\rm c2}$ curve is associated with the crossing of the metamagnetic transition at $H_{\rm x}$.
If there is no doubt on the strong ``S''-shaped curvature of $H_{\rm c2}(T)$,
the open question will be whether this phenomena is characteristic of FM2 phase 
or whether it is a combined effect of the volume expansion in the FM1 phase through the positive metamagnetic feedback,
leading to an increase of $T_{\rm sc}(P)$ just right at the maximum at $P_{\rm x}$.

The de Haas-van Alphen (dHvA) experiments under pressure reveal that
Fermi surfaces between FM2, FM1 and PM are quite different each other.
For $H \parallel b$-axis, where FM2, FM1 and PM phases are not affected by the magnetic fields,
dHvA branches of FM2 phase disappears in FM1 and new branches exhibit in FM2.~\cite{Ter01,Set02}
In PM phase, completely new branches are observed again.
The cyclotron effective mass gradually increases with increasing pressure up to $P_{\rm x}$
and in PM phase quite large effective masses ranging from $20$ to $60$ are detected,
in agreement with the pressure dependence of the $\gamma$-value.~\cite{Tat01}
For $H \parallel a$-axis, the field re-entrant FM1 and FM2 phases occurs, as shown in Fig.~\ref{fig:UGe2_TPH},
thus the results are more complicated.~\cite{Hag02,Ter02}
Nevertheless, the drastic change of Fermi surfaces is detected by crossing FM1, FM2 and PM phase boundary.
The cyclotron effective mass increases, approaching to $H_x$.
The change of Fermi surface is also found for $H\parallel c$-axis, as well.~\cite{Set03}

An interesting theoretical scenario proposed for the weak itinerant ferromagnet ZrZn$_2$ 
is the quantum metamagnetic transition associated with the 
topological change of Fermi surfaces, as proposed for a Lifshitz transition.~\cite{Yam07_Lifshitz}
The Fermi surface study under pressure can be found in Ref.~\citen{Kim04}.
In UGe$_2$, there is an evidence by the combined resistivity and Hall effect measurements~\cite{Kot11} that the
topological change of Fermi surface from PM to FM1 can be tuned by the pressure and field,
following the wing-shaped ($T,P,H$) phase diagram predicted in ``conventional'' spin fluctuation approaches of FM-QCEP.

\section{URhGe: a ferromagnetic superconductor at ambient pressure and field-reentrant SC}
Although the discovery of pressure induced SC in UGe$_2$ can be a major breakthrough,
the ambient pressure case provides much variety of experimental methods which goes deep inside the understanding of unconventional SC.
The discovery of SC at ambient pressure in the weak ferromagnet URhGe with $T_{\rm sc}=0.26\,{\rm K}$, $T_{\rm Curie}=9.5\,{\rm K}$ and $M_0=0.4\,\mu_{\rm B}$
opened the new opportunities.~\cite{Aok01}

\begin{figure}[tbh]
\begin{center}
\includegraphics[width=1 \hsize,clip]{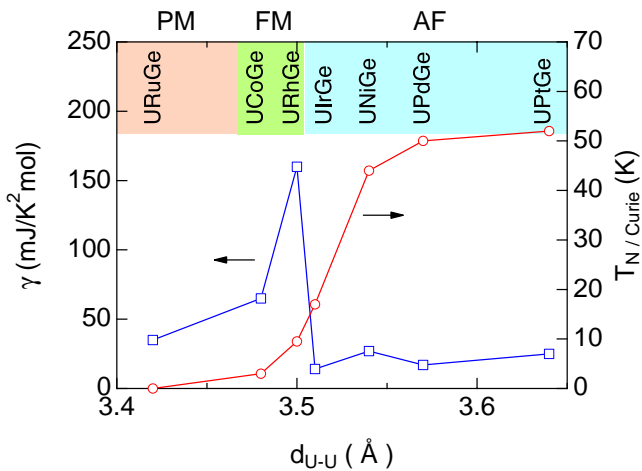}
\end{center}
\caption{(Color online) $\gamma$-value of specific heat and the magnetic ordered temperature as a function of the distance of the next nearest neighbor on U atom for UTGe (T: transition element) family. URuGe is paramagnet, UCoGe and URhGe are ferromagnets, the other UTGe are antiferromagnets.}
\label{fig:UTGe_Cp_distance}
\end{figure}

The properties of URhGe is summarized in Table~\ref{tab:table1}.
The crystal structure is orthorhombic TiNiSi-type, as shown in Fig.~\ref{fig:Structure}.
The U atom forms the zig-zag chain along $a$-axis with the distance of $d_{\mbox{U-U}}=3.50\,{\rm \AA}$,
which is close to the so-called Hill limit associated with the direct overlap of 5$f$-wave function.~\cite{Hil70}
Figure~\ref{fig:UTGe_Cp_distance} shows the $\gamma$-value and the magnetic ordered temperature as a function of the distance of 
the next nearest neighbor on U atom $d_{\mbox{U-U}}$ in UTGe (T: transition element) family.~\cite{Tro88}
The systematic variation can be seen. 
The PM ground state is realized for the small $d_{\mbox{U-U}}$, while the large $d_{\mbox{U-U}}$ induces the magnetic order with the large moment.
URhGe as well as UCoGe are located on the boundary between PM and AF with large moments accompanied with long range magnetic ordering.
The maximum and moderately enhanced $\gamma$-values are observed in URhGe and UCoGe.
The similar trend is also known in UX$_3$ and NpX$_3$ (X: group 13 and 14 elements).~\cite{Onu04,Aok06_NpIn3}

\begin{figure}[tbh]
\begin{center}
\includegraphics[width=1 \hsize,clip]{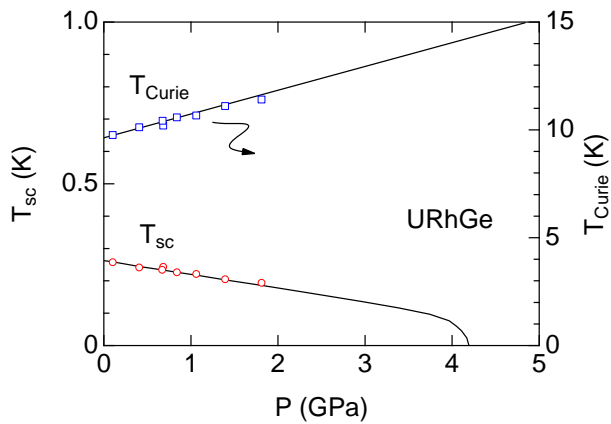}
\end{center}
\caption{(Color online) Pressure dependence of $T_{\rm Curie}$ and $T_{\rm sc}$ in URhGe.~\protect\cite{Har05_pressure,Miy09}}
\label{fig:URhGe_TP}
\end{figure}

In URhGe, $T_{\rm Curie}$ is much lower than than the band width, $T_{\rm Curie} \ll W$.
In the specific heat measurements, 
it is estimated that the contribution to the $\gamma$-value from the fit above $T_{\rm Curie}$ amounts to $\gamma_{\rm B}\sim 110\,{\rm mJ/K^2 mol}$,
while the FM fluctuation contributes $\sim 50\,{\rm mJ/K^2 mol}$ to the total $\gamma$-value ($160\,{\rm mJ/K^2 mol}$).~\cite{Har11}
Finally at low temperatures the relatively enhanced $\gamma$-value $160\,{\rm mJ/K^2mol}$ is achieved.
Contrary to UGe$_2$, with increasing pressure, $T_{\rm Curie}$ increases monotonously at least up to $12\,{\rm GPa}$, as shown in Fig.~\ref{fig:URhGe_TP},
indicating the system is pushed far from the FM instability.~\cite{Har05_pressure,Miy09}
Correspondingly the positive $\partial T_{\rm Curie}/\partial P$ is obtained from the Ehrenfest relation.~\cite{Har_pub}
The decrease of $T_{\rm sc}$ with pressure is associated with the decrease of $m^{\ast\ast}$,
which also implies that the system is getting away from the FM instability.

\begin{figure}[tbh]
\begin{center}
\includegraphics[width=1 \hsize,clip]{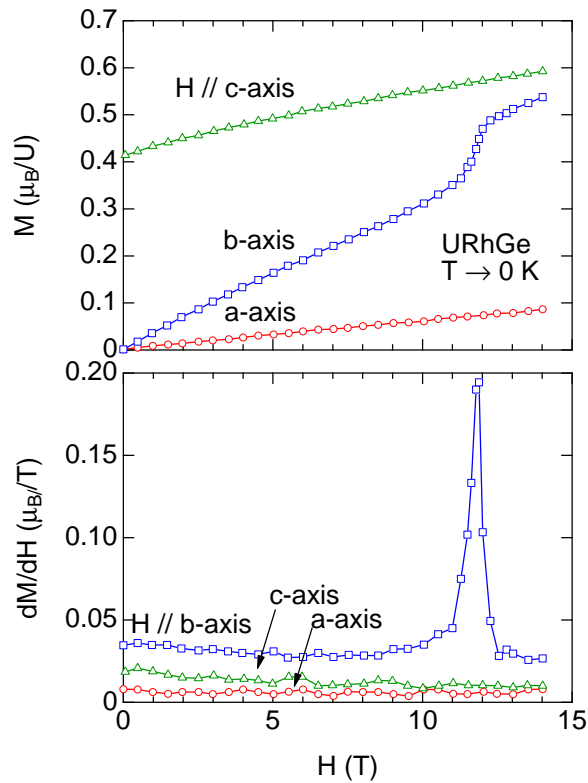}
\end{center}
\caption{(Color online) Magnetization curves and field derivative of magnetization in URhGe.~\protect\cite{Har11}}
\label{fig:URhGe_mag}
\end{figure}

The new feature is that the slope of magnetization curve $\partial M/\partial H$ 
for the field along the hard magnetization axis ($b$-axis) is larger than that along the easy axis ($c$-axis),
leading to the spin reorientation at $H_{\rm R}\sim 12\,{\rm T}$ for $H \parallel b$-axis, as shown in Fig.~\ref{fig:URhGe_mag}.
The magnetization curve for $H\parallel b$-axis displays that
the extrapolation of $M(H)$ from $H>H_{\rm R}$ to $H=0$ gives a finite value, ($\sim 0.15\,\mu_{\rm B}$).
This confirms that the system remains in FM side with the decrease of $M_0$ from $0.5\,\mu_{\rm B}$ to $0.15\,\mu_{\rm B}$,
as if the field sweep along $b$-axis leads to approach the FM instability.

\begin{figure}[tbh]
\begin{center}
\includegraphics[width=1 \hsize,clip]{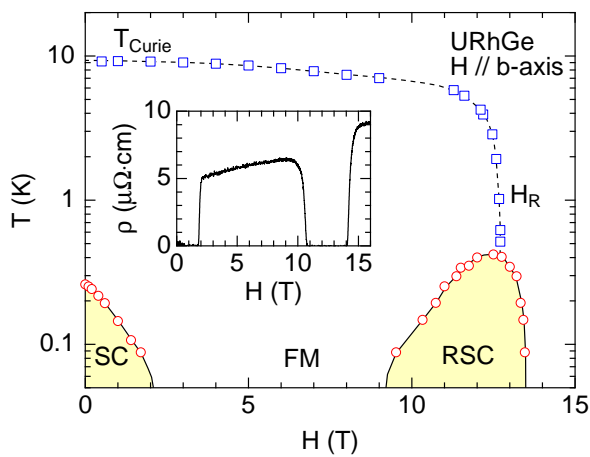}
\end{center}
\caption{(Color online) Temperature-field phase diagram for $H\parallel b$-axis in URhGe. SC, RSC and FM denote superconductivity, reentrant superconductivity and ferromagnetism, respectively. The inset shows the field dependence of resistivity at low temperatures ($\approx 80\,{\rm mK}$).~\protect\cite{Miy08} It is noted that the field range of RSC is very sensitive against the small mis-orientation to $c$-axis.}
\label{fig:URhGe_TH_resist}
\end{figure}

This spin reorientation gives rise to the field reentrant SC (RSC) around $H_{\rm R}$ at low temperatures.~\cite{Lev05}
Figure~\ref{fig:URhGe_TH_resist} shows the temperature-field phase diagram for $H\parallel b$-axis in URhGe.~\cite{Miy08}
Applying field, SC is suppressed around $2\,{\rm T}$, further increasing field, 
RSC appears approximately between $11\,{\rm T}$ and $14\,{\rm T}$.
Interestingly the maximum of $T_{\rm sc}$ for RSC phase ($\approx 0.42\,{\rm K}$)
is higher than $T_{\rm sc}$ for low field SC phase ($\approx 0.26\,{\rm K}$).
At high temperatures, $T_{\rm Curie}$ decreases with increasing fields 
as it is phenomenologically described by means of Landau free energy~\cite{Min11}.
The reduced $T_{\rm Curie}$ is connected to the spin reorientation field $H_{\rm R}$ at low temperatures.
The RSC as well as low field SC are very sensitive to the sample quality, 
indicating that both SCs are unconventional.~\cite{Miy08}
When the field is slightly tilted to the magnetization easy-axis ($c$-axis),
the RSC phase immediately shifts to higher fields and collapses.~\cite{Lev07,Aok11_ICHE}
This is attributed to the rapid suppression of longitudinal magnetic fluctuation by tilting field.
On the other hand, RSC is very robust when the field is tilted from $b$ to $a$-axis, i.e. maintaining the hard-magnetization axis.
$H_{\rm R}$ increases as a function of $1/\cos\theta$, where $\theta$ is the field angle from $b$ to $a$-axis.
Accompanying with the increase of $H_{\rm R}$, RSC is sustained even above $28\,{\rm T}$.~\cite{Lev07}

\begin{figure}[tbh]
\begin{center}
\includegraphics[width=1 \hsize,clip]{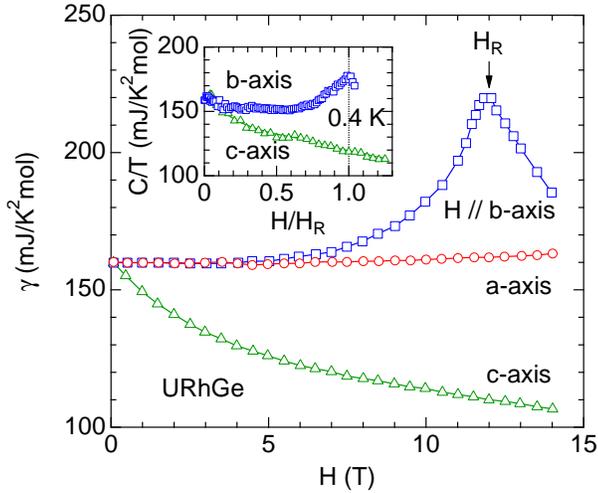}
\end{center}
\caption{(Color online) Field dependence of $\gamma$-values obtained by magnetization measurements via Maxwell relation. The initial $\gamma$-value at $0\,{\rm T}$ is taken as $160\,{\rm mJ/K^2mol}$.~\protect\cite{Har11} The inset shows the results of direct specific heat measurements at $0.4\,{\rm K}$.~\protect\cite{Aok11_ICHE} It is noted that the spin-reorientation field $H_{\rm R}$ on specific heat measurements slightly shifts to $15.2\,{\rm T}$, because of the small mis-orientation to $c$-axis within two degrees.}
\label{fig:URhGe_gamma}
\end{figure}

The origin of RSC is the enhancement of effective mass $m^\ast$ around $H_{\rm R}$
which is ascribed by approaching FM instability when a transverse field is applied in this Ising ferromagnets~\cite{Miy08,Aok11_pub,Har11}.
Figure~\ref{fig:URhGe_gamma} shows the field dependence of $\gamma$-value
obtained by the thermodynamic Maxwell relation via temperature dependence of the magnetization, i.e. $\partial \gamma/\partial H = \partial^2 M/ \partial T^2$.
The validity of this analysis via Maxwell relation was already demonstrated for CeRu$_2$Si$_2$~\cite{Pau90,Aok11_CeRu2Si2} and CeCoIn$_5$~\cite{Pau_pub}.
The inset shows the results of direct specific heat measurements for $H\parallel b$ and $c$-axis at $0.4\,{\rm K}$.
The similar results were also obtained by the field dependence of $A$ coefficient of $T^2$-term of resistivity
based on the so-called Kadowaki-Woods relation.~\cite{Miy08,Aok11_ICHE}.
The $\gamma$-value is enhanced around $H_{\rm R}$ for $H\parallel b$-axis, while it is suppressed for $H\parallel c$-axis,
indicating that the Ising-type FM fluctuation is enhanced around $H_{\rm R}$ for $H\parallel b$-axis.

The simple picture is that SC is related to the effective mass of conduction electron $m^\ast$
which has two major contributions: the renormalized band mass $m_{\rm B}$ and the effective mass gained through ferromagnetic or antiferromagnetic correlations $m^{\ast\ast}$, namely
\begin{equation}
m^\ast = m_{\rm B} + m^{\ast\ast}.
\end{equation}
Using McMillan-like formula,
the superconducting transition temperature $T_{\rm sc}$ is described by
\begin{equation}
T_{\rm sc}=T_0 \exp(-m^\ast/m^{\ast\ast}),
\end{equation}
where $T_0$ is the same as characteristic cut off energy.
Since $m^{\ast\ast}$ is strongly enhanced around $H_{\rm R}$, $T_{\rm sc}$ under fields is enhanced as well.
In ferromagnetic superconductors, the formation of spin-triplet state with equal-spin pairing is realized. 
$H_{\rm c2}$ is free from the Pauli limit based on the spin-singlet state, instead $H_{\rm c2}$ is governed by
the orbital limit $H_{\rm orb}$.
Since the superconducting coherence length $\xi$ is described by $\xi \approx \hbar v_{\rm F}^{}/k_{\rm B}T_{\rm sc}$,
we obtain a simple relation as $H_{\rm orb}\sim (m^\ast T_{\rm sc})^2$,
where $v_{\rm F}^{}$ is Fermi velocity.
If $m^\ast$ is enhanced, both $T_{\rm sc}$ and $H_{\rm orb}$ increase, and consequently RSC is observed at high fields.
It should be noted that enhancement of $T_{\rm sc}$ is affected by $m_{\rm B}$ and the Fermi surface.
To date, no drastic change of Fermi surface is inferred at $H_{\rm R}$ by thermopower measurements which will be published elsewhere.~\cite{Mal_pub}

\begin{figure}[tbh]
\begin{center}
\includegraphics[width=1 \hsize,clip]{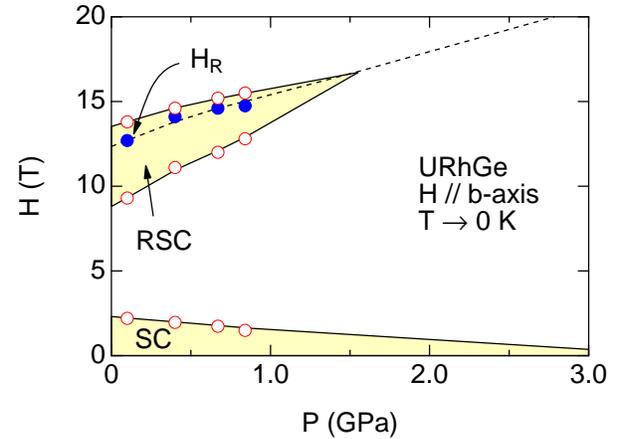}
\end{center}
\caption{(Color online) Pressure dependence of $H_{\rm c2}$ for low field SC and the critical fields for RSC for $H\parallel b$-axis in URhGe. $H_{\rm R}$ denotes the spin reorientation field.~\protect\cite{Miy09}}
\label{fig:URhGe_SC_pressure}
\end{figure}

Applying pressure, RSC phase shifts to higher fields associated with the increase of $H_{\rm R}$,
and eventually disappears above $\sim 1.5\,{\rm GPa}$, as shown in Fig.~\ref{fig:URhGe_SC_pressure}.~\cite{Miy09}
On the other hand, low-field SC will survive above $3\,{\rm GPa}$.
The suppression of both RSC and low-field SC is explained by the decrease of mass enhancement,
Interestingly, the similar behavior is observed at ambient pressure when the field is tilted to $c$-axis,~\cite{Aok11_ICHE}
where the longitudinal magnetic fluctuation is suppressed due to the Ising-type ferromagnetism.

At ambient pressure, $H_{\rm c2}$ of low field SC exceeds the Pauli limit for all three direction.
From the anisotropy of $H_{\rm c2}$, the line node gap in the $bc$-plane is inferred,
assuming the equal-spin pairing.~\cite{Har05},
where the attractive interaction between $\uparrow\uparrow$ electrons are described by
$V_{\sigma\sigma^\prime} (k,k^\prime) \propto \delta_{\uparrow\sigma}\delta_{\uparrow\sigma^\prime} k_a k_a^\prime$,
corresponding to an order parameter
$k_a | \uparrow\uparrow \rangle$.~\cite{Har05,Sch85}
It was demonstrated that this order parameter will remain when the FM moment rotates in the $bc$ plane.~\cite{Har05,Min06}

\section{UCoGe: strong interplay between FM and SC}
A new breakthrough was given by the discovery that the weak itinerant ferromagnet UCoGe 
($T_{\rm Curie}\sim 3\,{\rm K}$, $M_0 \sim 0.05\,\mu_{\rm B}$) becomes SC at $T_{\rm sc}\sim 0.7\,{\rm K}$.~\cite{Huy07}
The $\gamma$-value is moderately enhanced as $55\,{\rm mJ/K^2mol}$
The crystal structure is identical to that of URhGe with orthorhombic TiNiSi-type, as shown in Fig.~\ref{fig:Structure}
The distance of next nearest neighbor of U atom is slightly smaller than that of URhGe.
The magnetic moment is directed along $c$-axis, which is also identical to URhGe,
however the ordered moment ($\sim 0.05\,\mu_{\rm B}$) is much smaller that of URhGe.
At high fields for $H\parallel c$-axis, the magnetic moment is induced on the Co site with antiparallel direction.~\cite{Pro10}
The characteristic properties are summarized in Table~\ref{tab:table1}.
According to the band calculations based on the 5$f$-itinerant model with/without spin polarization,
the carrier number in PM state is small with semi-metallic type Fermi surface,
while the carrier number increases in FM state, but is still small.~\cite{Sam10}
The Shubnikov-de Haas experiments were carried out 
and a small pocket Fermi surface ($F\sim 1\,{\rm kT}$) was detected with large cyclotron mass ($25\,m_0$)~\cite{Aok11_UCoGe},
implying that UCoGe is a low carrier system with heavy quasi-particles.
This is also supported by the large Seebeck coefficient.~\cite{Mal_pub}
These situations are resemble to those of well-known semi-metallic heavy fermion superconductor URu$_2$Si$_2$.~\cite{Ohk99,Has10_URu2Si2,Zhu09,Shi09,Mal11}

\begin{figure}[tbh]
\begin{center}
\includegraphics[width=.8 \hsize,clip]{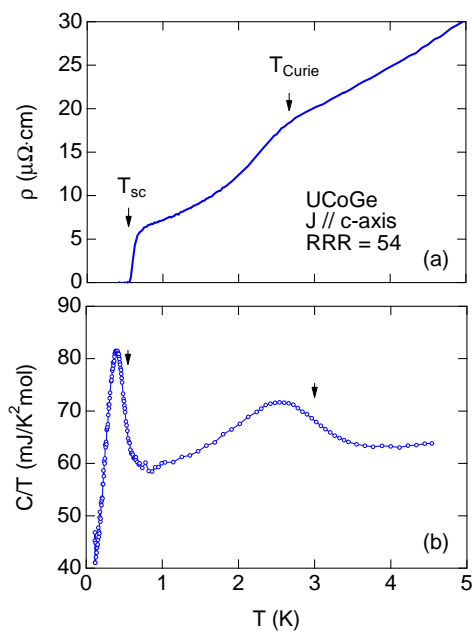}
\end{center}
\caption{(Color online) Temperature dependence of resistivity and specific heat in UCoGe.}
\label{fig:UCoGe_resist_Cp}
\end{figure}

In UCoGe the interplay between FM and SC is strong since $T_{\rm Curie}$ is already close to $T_{\rm sc}$.
Figure~\ref{fig:UCoGe_resist_Cp} represents the results of resistivity and specific heat.
Two anomalies are clearly observed at $T_{\rm Curie}$ and $T_{\rm sc}$ both in resistivity and in specific heat.
Contrary to URhGe, $T_{\rm Curie}$ is sensitive to the sample quality, while $T_{\rm sc}$ can detected for the poor quality sample with the residual resistivity ratio $\mbox{RRR}=3$ at least by resistivity measurements. 
This might be related with the fact that $T_{\rm Curie}$ is of first order and SC survives even above $P_{\rm c}$ in PM state, as described below.

\begin{figure}[tbh]
\begin{center}
\includegraphics[width=1 \hsize,clip]{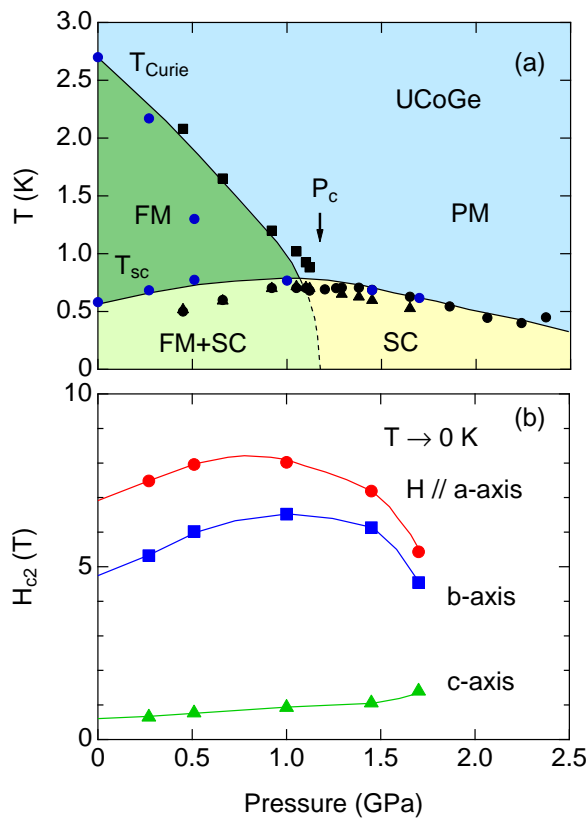}
\end{center}
\caption{(Color online) (a)Temperature--pressure phase diagram~\protect\cite{Has08,Slo09,Has10} and (b)pressure dependence of $H_{\rm c2}$ extrapolated to $0\,{\rm K}$ obtained by resistivity measurements down to $90\,{\rm mK}$ in UCoGe. It is noted that the values of $H_{\rm c2}$ for $a$ and $b$-axis are reduced by comparison to those of perfectly aligned field direction. The mis-orientation to $c$-axis is estimated to be within two degrees.}
\label{fig:UCoGe_TP_phase}
\end{figure}

Applying the pressure, $T_{\rm Curie}$ and $T_{\rm sc}$ are merged around the critical pressure $P_{\rm c}\sim 1\,{\rm GPa}$.~\cite{Has08,Slo09,Has10}
AC susceptibility and AC calorimetry measurements clearly established that SC 
is very robust through $P_{\rm c}$.
The simple idea is that when $T_{\rm Curie}$ is close to  $T_{\rm sc}$, 
FM collapses via the first order transition
and the associated volume discontinuity gives rise to the system with comparable FM fluctuations and thus
to comparable $T_{\rm sc}$ on both sides of $P_{\rm c}$.
The phase diagram shown in Fig.~\ref{fig:UCoGe_TP_phase}(a) is contradictory to the theoretical prediction near FM fluctuation by Fay and Appel,~\cite{Fay80}
assuming a second order quantum critical point.~\cite{Fay80}
A new theory through symmetry approach has been proposed for the ($T,P$) phase diagram of UCoGe.~\cite{Min08}
Experimentally the domain of coexistence between SC and FM with $T_{\rm sc} < T_{\rm Curie}$ has not been detected.

Pressure dependence of $H_{\rm c2}$ extrapolated to $0\,{\rm K}$ is shown in Fig.~\ref{fig:UCoGe_TP_phase}(b).
It should be noted that the values of $H_{\rm c2}$ are reduced, 
compared to those for the perfectly field-aligned case,
since the values of $H_{\rm c2}$ for $a$ and $b$-axis are very sensitive 
to the field mis-orientation to $c$-axis, as mentioned later.
Nevertheless, measured $H_{\rm c2}$ curves for all field direction reveal almost linear increase with slight upward curvature with decreasing temperature for all the pressure range (not shown),
thus we can determine the pressure dependence of $H_{\rm c2}$ at $0\,{\rm K}$.
For $a$ and $b$-axis, broad maxima of $H_{\rm c2}$ are observed around $P_{\rm c}$,
associated with the broad maximum of $T_{\rm sc}$.
On the other hand, $H_{\rm c2}$ for $H \parallel c$-axis increases monotonously with pressure,
indicating that the anisotropies between $a$ and $c$ or between $b$ and $c$, i.e. $H_{\rm c2}^a/H_{\rm c2}^c$ and $H_{\rm c2}^b/H_{\rm c2}^c$
are reduced above $P_{\rm c}$.
This may be due to the fact that longitudinal magnetic fluctuation is suppressed above $P_{\rm c}$.
The first order nature of $T_{\rm Curie}$ and the acute $H_{\rm c2}$ enhancement for $H\parallel a$ and $b$-axis impede 
the precise determination of $H_{\rm c2}$ as a function of pressure at the moment.
Another supplement event might be the change of Fermi surface at the FM--PM transition.

\begin{figure}[tbh]
\begin{center}
\includegraphics[width=1 \hsize,clip]{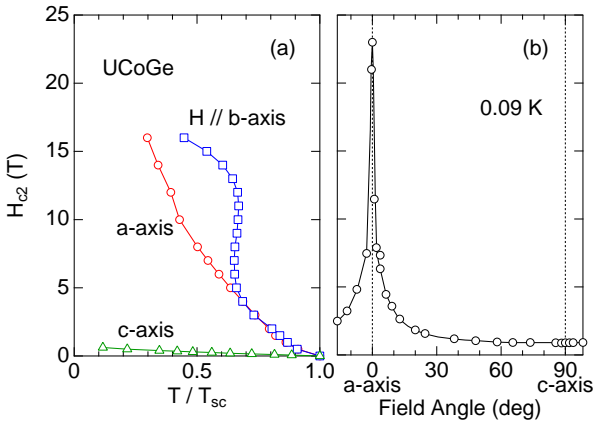}
\end{center}
\caption{(Color online) (a)Temperature dependence of $H_{\rm c2}$ and (b)angular dependence of $H_{\rm c2}$ from $a$ to $c$-axis at $0.09\,{\rm K}$ in UCoGe. $T_{\rm sc}$ is $\sim 0.65\,{\rm K}$ at $0\,{\rm T}$ .~\protect\cite{Aok09_UCoGe}}
\label{fig:UCoGe_Hc2}
\end{figure}

The new striking point is that $H_{\rm c2}$ at ambient pressure is strongly enhanced
when the field is applied along the hard magnetization axis ($a$ and $b$-axis), as shown in Fig.~\ref{fig:UCoGe_Hc2}(a).~\cite{Aok09_UCoGe,Aok11_ICHE}
At $0,{\rm K}$, $H_{\rm c2}$ for $H\parallel b$ and $a$-axis reaches $H_{\rm c2}^b \sim 18\,{\rm T}$ and $H_{\rm c2}^a > 30\,{\rm T}$,
which considerably exceed the Pauli limit estimated from $T_{\rm sc}\sim 0.6\,{\rm K}$.
On the other hand, $H_{\rm c2}$ for $H \parallel c$-axis is $0.6\,{\rm T}$, which is comparable or even less than the Pauli limit.
The acute enhanced $H_{\rm c2}$ can be seen for $H\parallel a$-axis, as shown in Fig.~\ref{fig:UCoGe_Hc2}(b).
The fact that $H_{\rm c2}$ is strongly damped by tilting the field angle slightly to $c$-axis cannot be explained by the 
conventional effective mass model associated with the Fermi surface topology, but
should be ascribed by the anisotropic magnetic fluctuation. 
The huge anisotropy of $H_{\rm c2}$ including an ``S''-shaped curve for $b$-axis is
qualitatively explained by the anisotropic field response of effective mass.
The field dependence of $A$ coefficient of $T^2$ term of resistivity, which is linked to the $\gamma$-value and effective mass by Kadowaki-Woods relation assuming the strong local fluctuation ($A\propto \gamma^2 \propto {m^\ast}^2$),
shows that $A$ for $H\parallel c$-axis is suppressed with field as usual weak itinerant ferromagnets,
while $A$ for $H\parallel b$ and $a$-axis remains at high value, in addition, $A$ for $H\parallel b$-axis reveals the maximum at field
where the ``S''-shaped $H_{\rm c2}$ is observed.
The results are similar to those obtained in URhGe, as shown in Fig.~\ref{fig:URhGe_gamma}.

NMR and NQR experiments are very good probes to study magnetic fluctuations
associated with dynamic as well as static susceptibilities, 
because $^{59}$Co is an excellent nuclear in UCoGe.
An evidence for the first order transition at $T_{\rm Curie}$ at ambient pressure was given from the abrupt change of 
the resonant frequency below $T_{\rm Curie}$.~\cite{Oht10}
This implies that UCoGe is already above the tricritical point.
The nuclear spin-lattice relaxation rate $1/T_1$ measurements microscopically confirm 
the coexistence of FM and SC, and the strong Ising character of the ferromagnetic fluctuation.~\cite{Oht10}
The dynamic susceptibility shows the remarkable anisotropy, implying 
that the longitudinal FM magnetic fluctuation is dominated.~\cite{Iha10}

The results of experiments on three ferromagnetic superconductors, UGe$_2$, URhGe and UCoGe
confirm that Ising-type FM with longitudinal fluctuation mode is favorable for SC,
Up to now, there are no other cases of coexistence of FM and SC despite the attempt 
to find Ce based heavy fermion compounds.
Let us remark that SC near AF criticality seems to be favored by transverse spin fluctuations 
as it occurs in CeCu$_2$Si$_2$, CePd$_2$Si$_2$, Ce and Pu-115 systems and NpPd$_5$Al$_2$~\cite{Aok07_NpPd5Al2,Chu10,Kam07b,Miy86,Sca86},
while for Ising-type AF system, such as CeRu$_2$Si$_2$, no SC has been observed.

In contrast to URhGe, no spin reorientation is expected in UCoGe for $H\parallel b$-axis as the slope of magnetization $\chi_b$
is smaller than $\chi_c$.
The key criteria will be the size of the field induced magnetization $M_b$
by comparison to the ordered moment $M_0$ at zero field.
When the magnetic field reaches $H_b$ where $M_b$ is comparable to $M_0$, 
namely $M_0 \approx M_b = \chi_b H_b$,
the drastic change of effective mass is expected.
Table~\ref{tab:table2} summarizes the parameters of three uranium ferromagnetic superconductors.
\begin{table}[tbhp]
\caption{Susceptibilities and characteristic fields of UGe$_2$, URhGe and UCoGe.}
\begin{center}
\begin{tabular}{ccccccc}
\hline
			& $\chi_a$	& $\chi_b$	& $\chi_c$			& $H_a$		& $H_b$		& $H_c$		\\
			& \multicolumn{3}{c}{($\mu_{\rm B}/{\rm T}$)}	& \multicolumn{3}{c}{(T)}		\\
\hline
UGe$_2$		& 0.006		& 0.0055		& 0.011				& 230		& 250		& 122		\\
URhGe		& 0.006		& 0.03		& 0.01				& 66			& 13			& 40			\\
UCoGe		& 0.0024		& 0.006		& 0.029				& 29			& 12			& 2.5		\\
\hline
\end{tabular}
\end{center}
\label{tab:table2}
\end{table}%

An interesting unique point in ferromagnetic superconductors is that 
the internal field associated with the ordered moment $M_0$ is large compared to the expected value of superconducting lower critical field $H_{\rm c1}$.
Thus spontaneous vortex state must be realized at zero field.
So far there is no direct observation of the corresponding vortex lattice, but clear marks
are obtained by NQR measurements in UCoGe~\cite{Oht10} 
and the unusual initial slope of $H_{\rm c2}$ in URhGe.~\cite{Har05}

\begin{figure}[tbh]
\begin{center}
\includegraphics[width=1 \hsize,clip]{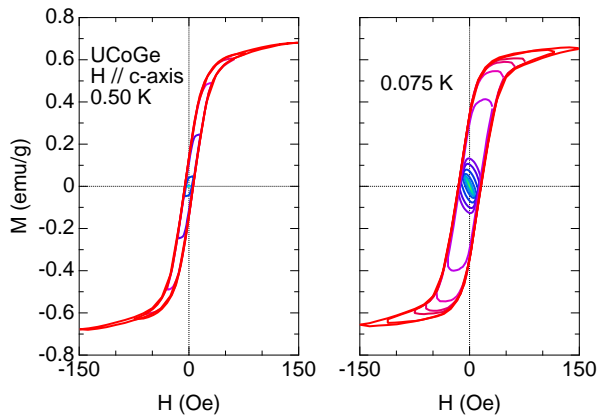}
\end{center}
\caption{(Color online) Hysteresis loops of magnetizations under various field-sweep windows above and below $T_{\rm sc}$ for $H\parallel c$-axis in UCoGe.~\protect\cite{Pau_pub}}
\label{fig:UCoGe_hysteresis}
\end{figure}

An interesting macroscopic observation is the modification of the hysteresis loop of magnetization through the SC transition,
as shown in Fig.~\ref{fig:UCoGe_hysteresis}.~\cite{Pau_pub}
Careful analysis of the DC magnetization shows that no $H_{\rm c1}$ exists at least for $H \parallel c$-axis.
The similar experiments can be found in Ref.~\citen{Deg10}.

\section{Theoretical view}
The first prediction of triplet SC in metallic system near FM instability was reported by Fay and Appel~\cite{Fay80}.
$T_{\rm sc}$ reveals two maxima both in PM and in FM phase.
The very slow energy fluctuation gives rise to the pair-breaking in the vicinity of $P_{\rm c}$,
while $p$-wave SC in nearly FM itinerant system was already calculated by Layzer and Fay in 1971.~\cite{Lay71}
Recent discussion for nearly ferromagnetic system can be found in Ref.~\citen{Mon01,Mon07,Fuj04}.
A more complicated treatment wipes out the collapse of $T_{\rm sc}$ at $P_{\rm c}$, instead
only a minimum of $T_{\rm sc}$ will occur.
However, the experimental observation for UCoGe is that $T_{\rm sc}$ has a broad maximum at $P_{\rm c}$
accompanied with the first order transition of $T_{\rm Curie}$, instead of second order (see Fig.~\ref{fig:UCoGe_TP_phase}).
Discussion on the coexistence of SC and FM can be found in Ref.~\citen{Kir03}
An interesting point is also that in the FM state $T_{\rm sc}$ for the majority spin ($\uparrow$)
differs from that for the minority spin ($\downarrow$): $T_{\downarrow\downarrow} > T_{\uparrow\uparrow}$.
Thus SC in the FM region is a two-band superconductor with the possibility that only one type of band is gapped.

The possible order parameters in FM phase have been classified on general symmetry arguments~\cite{Sam02,Fom01,Min02}.
On the basis of the report of SC in the cubic ZrZn$_2$~\cite{Pfl01}, which was found to be extrinsic afterward~\cite{Yel05},
it was predicted that the gap nodes will change when the magnetization is rotated by magnetic field.~\cite{Wal02}
It was also proposed that in the weak Heisenberg ferromagnet $T_{\rm sc}$ will be enhanced on the FM side
due to the development of transverse magnetic fluctuation.
To date, the evidences of SC in FM materials is limited to Ising-type FM.

For UGe$_2$, the striking point is that $T_{\rm sc}$ has maximum at $P_{\rm x}$.
It was proposed that the CDW/SDW fluctuation may occur~\cite{Wat02}, 
but up to now no extra superstructure was found experimentally.
Phenomenological model was developed assuming a twin-peak in the electronic density of states.~\cite{San03}
It was even proposed that for UGe$_2$ close to $P_{\rm x}$ as $M_0$ is still high and thus
the magnetism is based on the strongly localized case that the coupling of two electrons via localized spin
can be attractive~\cite{Suh01},
and demonstrated that this $s$-wave attraction holds for the whole Fermi surface.~\cite{Abr01}
However, this hypothesis is questionable 
because the Fermi surfaces calculations clearly show that
5$f$ electron must be considered as itinerant and 
this character is strongly reinforced in FM1 phase.

Interesting new features are predicted as the SC order parameter is linked to $M_0$
with the possibility that domain walls play a role of weak links.~\cite{Fom01,Mac01}
It was also stressed that SC may appear locally at a domain wall not inside a magnetic domain.~\cite{Buz03}

Recently it was stressed in good agreement with the study made on URhGe that
when the field is applied along $b$-axis ($\perp M_0$), 
$T_{\rm Curie}(H)$ will decrease in a quadratic dependence ($\Delta T_{\rm Curie} \propto -H^2$)~\cite{Min11}.
Analysis of the unusual temperature dependence and the anisotropy of $H_{\rm c2}(T)$ was made close to $P_{\rm c}$.
Due to the long range nature of the FM interaction,
non-analytic correction can enhance the SC transition.~\cite{Tad_ICHE}
The position of nodes with respect to the magnetic field direction 
can explain the unusual angular dependence of $H_{\rm c2}$.
For UCoGe it was proposed that the large anisotropy between $H_{\rm c2}$ for $H\parallel a$-axis ($H_{\rm c2}^a$) and $H_{\rm c2}^c$,
namely $H_{\rm c2}^a \gg H_{\rm c2}^c$ implies a point node gap, not an horizontal line node gap, 
with respect to the vertical $M_0$ direction.

\section{Conclusion}
Discovery of three uranium ferromagnetic superconductors has open the interesting frontiers 
for the interplay between two major ground states of condensed matter, FM and SC.
Experimentally, a key challenge is to discover an ideal simple case,
such as Ce-115 systems for the interplay of AF and SC,
where the high quality single crystals with large size can be obtained.

UGe$_2$ is unfortunately not an ideal example for SC despite the availability of high quality single crystal,
since the external pressure is required and there is not the sole transition from FM to PM,
but also the switch from FM1 to FM2.
Furthermore the changes of ground state occurs at marked first order transitions.
It is an excellent example to study the tricriticality and the properties at QCEP.

URhGe and UCoGe suffer from the unavailability of large and high quality single crystals.
Thus, careful tests are required to be sure that the measurements characterize the bulks homogeneous SC.
At least clear new phenomena have emerged 
such as the link for field-reentrant SC in transverse field response with respect to the easy magnetization axis in URhGe and UCoGe.

\section*{Acknowledgments}
We thank J. P. Brison, A. Buzdin, S. Fujimoto, Y. Haga, F. Hardy, H. Harima, K. Hasselbach, E. Hassinger, L. Howald, K. Ishida, S. Kambe, W. Knafo, G. Knebel, H. Kotegawa, L. Malone, T. D. Matsuda, C. Meingast, V. Michal, V. Mineev, A. Miyake, K. Miyake, C. Paulsen, S. Raymond, R. Settai, I. Sheikin, Y. Tada, V. Taufour, for fruitful discussion. 
This work was supported by ERC starting grant (NewHeavyFermion), French ANR project (CORMAT, SINUS, DELICE) and REIMEI program in JAEA.


\begin{thebibliography}{10}

\bibitem{Fis90_FM}
{\O}.~Fischer: {\em Magnetic Superconductors in Ferromagnetic Materials}
  (Science Publishers BV, Amsterdam, 1990).

\bibitem{Flo02}
J.~Flouquet and A.~Buzdin: Physics World (2002) 41.

\bibitem{Ish77}
M.~Ishikawa and O.~Fischer: Solid State Commun. {\bf 23} (1977) 37.

\bibitem{Fer77}
W.~A. Fertig, D.~C. Johnston, L.~E. DeLong, R.~W. McCallum, M.~B. Maple and
  B.~T. Matthias: Phys. Rev. Lett. {\bf 38} (1977) 987.

\bibitem{Jac62}
V.~Jaccarino and M.~Peter: Phys. Rev. Lett. {\bf 9} (1962) 290.

\bibitem{Meu84}
H.~W. Meul, C.~Rossel, M.~Decroux, {\O}.~Fischer, G.~Remenyi and A.~Briggs:
  Phys. Rev. Lett. {\bf 53}  (1984) 497.

\bibitem{Bal01}
L.~Balicas, J.~S. Brooks, K.~Storr, S.~Uji, M.~Tokumoto, H.~Tanaka,
  H.~Kobayashi, A.~Kobayashi, V.~Barzykin and L.~P. Gor'kov: Phys. Rev. Lett.
  {\bf 87}  (2001) 067002.

\bibitem{Uji01}
S.~Uji, H.~Shinagawa, T.~Terashima, T.~Yakabe, Y.~Terai, M.~Tokumoto,
  A.~Kobayashi, H.~Tanaka and H.~Kobayashi: Nature {\bf 410} (2001) 908.

\bibitem{Flo06_review}
J.~Flouquet: {\em Progress in Low Temperature Physics}, ed. W.~P. Halperin
  (Elsevier, Amsterdam, 2006) p.~139.

\bibitem{Leg75}
A.~J. Leggett: Rev. Mod. Phys. {\bf 47} (1975) 331.

\bibitem{Flo82}
J.~Flouquet, J.~C. Lasjaunias, J.~Peyrard and M.~Ribault: J. Appl. Phys. {\bf
  53} (1982) 2127.

\bibitem{Ste79}
F.~Steglich, J.~Aarts, C.~D. Bredl, W.~Lieke, D.~Meschede, W.~Franz and
  H.~{Sch\"{a}fer}: Phys. Rev. Lett. {\bf 43} (1979) 1892.

\bibitem{Sax00}
S.~S. Saxena, P.~Agarwal, K.~Ahilan, F.~M. Grosche, R.~K.~W. Haselwimmer, M.~J.
  Steiner, E.~Pugh, I.~R. Walker, S.~R. Julian, P.~Monthoux, G.~G. Lonzarich,
  A.~Huxley, I.~Sheikin, D.~Braithwaite and J.~Flouquet: Nature {\bf 406}
  (2000) 587.

\bibitem{Aok01}
D.~Aoki, A.~Huxley, E.~Ressouche, D.~Braithwaite, J.~Flouquet, J.-P. Brison,
  E.~Lhotel and C.~Paulsen: Nature {\bf 413} (2001) 613.

\bibitem{Huy07}
N.~T. Huy, A.~Gasparini, D.~E. {de Nijs}, Y.~Huang, J.~C.~P. Klaasse,
  T.~Gortenmulder, A.~{de Visser}, A.~Hamann, T.~{G\"{o}rlach} and
  H.~v.~{L\"{o}hneysen}: Phys. Rev. Lett. {\bf 99} (2007) 067006.

\bibitem{Pfl01}
C.~Pfleiderer, M.~Uhlarz, S.~M. Hayden, R.~Vollmer, H.~von L{\"o}hneysen, N.~R.
  Bernhoeft and G.~G. Lonzarich: NATURE {\bf 412} (2001) 58.

\bibitem{Yel05}
E.~A. Yelland, S.~M. Hayden, S.~J.~C. Yates, C.~Pfleiderer, M.~Uhlarz,
  R.~Vollmer, H.~v. L{\"o}hneysen, N.~R. Bernhoeft, R.~P. Smith, S.~S. Saxena
  and N.~Kimura: Phys. Rev. B {\bf 72} (2005) 214523.

\bibitem{Aok11_CR}
D.~Aoki, F.~Hardy, A.~Miyake, V.~Taufour, T.~D. Matsuda and J.~Flouquet: C. R.
  Physique {\bf 12} (2011) 573.

\bibitem{Oik96}
K.~Oikawa, T.~Kamiyama, H.~Asano, Y.~\={O}nuki and M.~Kohgi: J. Phys. Soc. Jpn.
  {\bf 65} (1996) 3229.

\bibitem{Onu92}
Y.~\={O}nuki, I.~Ukon, S.~W. Yun, I.~Umehara, K.~Satoh, T.~Fukuhara, H.~Sato,
  S.~Takayanagi, M.~Shikama and A.~Ochiai: J. Phys. Soc. Jpn. {\bf 61}
  (1992) 293.

\bibitem{Oom95}
G.~Oomi, T.~Kagayama, K.~Nishimura, S.~W. Yun and Y.~Onuki: Physica B:
  Condensed Matter {\bf 206-207} (1995) 515.

\bibitem{Tau11}
V.~Taufour, A.~Villaume, D.~Aoki, G.~Knebel and J.~Flouquet: J. Phys.: Conf.
  Ser. {\bf 273} (2011) 012017.

\bibitem{Hux01}
A.~Huxley, I.~Sheikin, E.~Ressouche, N.~Kernavanois, D.~Braithwaite,
  R.~Calemczuk and J.~Flouquet: Phys. Rev. B {\bf 63} (2001) 144519.

\bibitem{Hux00}
A.~Huxley, I.~Sheikin and D.~Braithwaite: Physica B {\bf 284-288} (2000) 1277.

\bibitem{Pfl02}
C.~Pfleiderer and A.~D. Huxley: Phys. Rev. Lett. {\bf 89} (2002) 147005.

\bibitem{Hux03}
A.~Huxley, E.~Ressouche, B.~Grenier, D.~Aoki, J.~Flouquet and C.~Pfleiderer: J.
  Phys.: Condens. Matter {\bf 15} (2003) S1945.

\bibitem{Kot05}
H.~Kotegawa, A.~Harada, S.~Kawasaki, Y.~Kawasaki, Y.~Kitaoka, Y.~Haga,
  E.~Yamamoto, Y.~\={O}nuki, K.~M. Itoh, E.~E. Haller and H.~Harima: J. Phys.
  Soc. Jpn. {\bf 74} (2005) 705.

\bibitem{Tat01}
N.~Tateiwa, T.~C. Kobayashi, K.~Hanazono, K.~Amaya, Y.~Haga, R.~Settai and
  Y.~\={O}nuki: J. Phys.: Condens. Matter {\bf 13} (2001) L17.

\bibitem{Tat04}
N.~Tateiwa, T.~C. Kobayashi, K.~Amaya, Y.~Haga, R.~Settai and Y.~\={O}nuki:
  Phys. Rev. B {\bf 69} (2004) 180513.

\bibitem{Tau10}
V.~Taufour, D.~Aoki, G.~Knebel and J.~Flouquet: Phys. Rev. Lett. {\bf 105}
  (2010) 217201.

\bibitem{Kab10}
N.~Kabeya, R.~Iijima, E.~Osaki, S.~Ban, K.~Imura, K.~Deguchi, N.~Aso, Y.~Homma,
  Y.~Shiokawa and N.~K. Sato: J. Phys.: Conf. Ser. {\bf 200} (2010) 032028.

\bibitem{Kot11} 
H. Kotegawa, V. Taufour, D. Aoki, G. Knebel and J. Flouquet: J. Phys. Soc. Jpn. {\bf 80} 083703.

\bibitem{She01}
I.~Sheikin, A.~Huxley, D.~Braithwaite, J.~P. Brison, S.~Watanabe, K.~Miyake and
  J.~Flouquet: Phys. Rev. B {\bf 64} (2001) 220503.

\bibitem{Ter01}
T.~Terashima, T.~Matsumoto, C.~Terakura, S.~Uji, N.~Kimura, M.~Endo,
  T.~Komatsubara and H.~Aoki: Phys. Rev. Lett. {\bf 87}  (2001) 166401.

\bibitem{Set02}
R.~Settai, M.~Nakashima, S.~Araki, Y.~Haga, T.~C. Kobayashi, N.~Tateiwa,
  H.~Yamagami and Y.~\={O}nuki: J. Phys.: Condens. Matter {\bf 14} (2002) L29.

\bibitem{Hag02}
Y.~Haga, M.~Nakashima, R.~Settai, S.~Ikeda, T.~Okubo, S.~Araki, T.~C. K.~N.
  Tateiwa and Y.~\={O}nuki: J. Phys.: Condens. Matter {\bf 14} (2002) L125.

\bibitem{Ter02}
T.~Terashima, T.~Matsumoto, C.~Terakura, S.~Uji, N.~Kimura, M.~Endo,
  T.~Komatsubara, H.~Aoki and K.~Maezawa: Phys. Rev. B {\bf 65} (2002) 174501.

\bibitem{Set03}
R.~Settai, M.~Nakashima, H.~Shishido, Y.~Haga, H.~Yamagami and Y.~\={O}nuki:
  Acta Physica Polonica B {\bf 34} (2003) 725.

\bibitem{Yam07_Lifshitz}
Y.~Yamaji, T.~Misawa and M.~Imada: J. Phys. Soc. Jpn. {\bf 76} (2007) 063702.

\bibitem{Kim04}
N.~Kimura, M.~Endo, T.~Isshiki, S.~Minagawa, A.~Ochiai, H.~Aoki, T.~Terashima,
  S.~Uji, T.~Matsumoto and G.~G. Lonzarich: Phys. Rev. Lett. {\bf 92} 
  (2004) 197002.

\bibitem{Hil70}
H.~H. Hill: {\em Plutonium and Other Actinides}, ed. W.~N. Miner (AIME, New
  York, 1970) p.~2.

\bibitem{Tro88}
R.~Tro{\'c} and V.~H. Tran: J. Magn. Magn. Mater. {\bf 73} (1988) 389.

\bibitem{Onu04}
Y.~\={O}nuki, R.~Settai, K.~Sugiyama, T.~Takeuchi, T.~C. Kobayashi, Y.~Haga and
  E.~Yamamoto: J. Phys. Soc. Jpn. {\bf 73}  (2004) 769.

\bibitem{Aok06_NpIn3}
D.~Aoki, Y.~Homma, H.~Sakai, S.~Ikeda, Y.~Shiokawa, E.~Yamamoto, A.~Nakamura,
  Y.~Haga, R.~Settai and Y.~\={O}nuki: J. Phys. Soc. Jpn. {\bf 74} (2006)
  084710.

\bibitem{Har11} 
F. Hardy, D. Aoki, C. Meingast, P. Schweiss, P. Burger, H. v. Loehneysen and J. Flouquet: 
Phys. Rev. B {\bf 83}, 195107 (2011).

\bibitem{Har05_pressure}
F.~Hardy, A.~Huxley, J.~Flouquet, B.~Salce, G.~Knebel, D.~Braithwaite, D.~Aoki,
  M.~Uhlarz and C.~Pfleiderer: Physica B {\bf 359} (2005) 1111.

\bibitem{Miy09}
A.~Miyake, D.~Aoki and J.~Flouquet: J. Phys. Soc. Jpn. {\bf 78} (2009) 063703.

\bibitem{Har_pub}
{F. Hardy et al.}: to be published.

\bibitem{Lev05}
F.~L\'{e}vy, I.~Sheikin, B.~Grenier and A.~D. Huxley: Science {\bf 309} (2005)
  1343.

\bibitem{Miy08}
A.~Miyake, D.~Aoki and J.~Flouquet: J. Phys. Soc. Jpn. {\bf 77} (2008) 094709.

\bibitem{Min11}
V.~P. Mineev: Phys. Rev. B {\bf 83}  (2011) 064515.

\bibitem{Lev07}
F.~L\'{e}vy, I.~Sheikin and A.~Huxley: Nature Physics {\bf 3} (2007) 460.

\bibitem{Aok11_ICHE}
D.~Aoki, T.~D. Matsuda, F.~Hardy, C.~Meingast, V.~Taufour, E.~Hassinger,
  I.~Sheikin, C.~Paulsen, G.~Knebel, H.~Kotegawa and J.~Flouquet: J. Phys. Soc.
  Jpn. {\bf 80} (2011) SA008.

\bibitem{Aok11_pub}
D.~Aoki, H.~Kotegawa and J.~Flouquet: {to be published}.

\bibitem{Pau90}
C.~Paulsen, A.~Lacerda, L.~Puech, P.~Haen, P.~Lejay, J.~L. Tholence,
  J.~Flouquet and A.~{de Visser}: J. Low Temp. Phys. {\bf 81} (1990) 317.

\bibitem{Aok11_CeRu2Si2}
D.~Aoki, C.~Paulsen, T.~D. Matsuda, L.~Malone, G.~Knebel, P.~Haen, P.~Lejay,
  R.~Settai, Y.~\={O}nuki and J.~Flouquet: J. Phys. Soc. Jpn. {\bf 80} (2011)
  053702.

\bibitem{Pau_pub}
C.~Paulsen, D.~Aoki, G.~Knebel and J.~Flouquet: arXiv:1102.2703.

\bibitem{Mal_pub}
L.~Malone, L.~Howald, A.~Pourret, D.~Aoki, V.~Taufour, G.~Knebel and J. Flouquet: to be published.

\bibitem{Har05}
F.~Hardy and A.~D. Huxley: Phys. Rev. Lett. {\bf 94} (2005) 247006.

\bibitem{Sch85}
K.~Scharnberg and R.~A. Klemm: Phys. Rev. Lett. {\bf 54} (1985) 2445.

\bibitem{Min06}
V.~P. Mineev: C. R. Physique {\bf 7} (2006) 35.

\bibitem{Pro10}
K.~Proke\v{s}, A.~{de Visser}, Y.~K. Huang, B.~F{\aa}k and E.~Ressouche: Phys.
  Rev. B {\bf 81} (2010) 180407.

\bibitem{Sam10}
M.~{Samsel-Czeka{\l}a}, S.~Elgazzar, P.~M. Oppeneer, E.~Talik, W.~Walerczyk and
  R.~Tro{\'c}: J. Phys.: Condens. Matter {\bf 22} (2010) 015503.

\bibitem{Aok11_UCoGe}
D.~Aoki, I.~Sheikin, T.~D. Matsuda, V.~Taufour, G.~Knebel and J.~Flouquet: J.
  Phys. Soc. Jpn. {\bf 80} (2011) 013705.

\bibitem{Ohk99}
H.~Ohkuni, Y.~Inada, Y.~Tokiwa, K.~Sakurai, R.~Settai, T.~Honma, Y.~Haga,
  E.~Yamamoto, Y.~\={O}nuki, H.~Yamagami, S.~Takahashi and T.~Yanagisawa:
  Philos. Mag. B {\bf 79} (1999) 1045.

\bibitem{Has10_URu2Si2}
E.~Hassinger, G.~Knebel, T.~D. Matsuda, D.~Aoki, V.~Taufour and J.~Flouquet:
  Phys. Rev. Lett. {\bf 105} (2010) 216409.

\bibitem{Zhu09}
Z.~Zhu, E.~Hassinger, Z.~Xu, D.~Aoki, J.~Flouquet and K.~Behnia: Phys. Rev. B
  {\bf 80} (2009) 172501.

\bibitem{Shi09}
H.~Shishido, K.~Hashimoto, T.~Shibauchi, T.~Sasaki, H.~Oizumi, N.~Kobayashi,
  T.~Takamasu, K.~Takehana, Y.~Imanaka, T.~D. Matsuda, Y.~Haga, Y.~Onuki and
  Y.~Matsuda: Phys. Rev. Lett. {\bf 102} (2009) 156403.

\bibitem{Mal11}
L.~Malone, T.~D. Matusda, A.~Antunes, G.~Knebel, V.~Taufour, D.~Aoki,
  K.~Behnia, C.~Proust and J.~Flouquet: Phys. Rev. B {\bf 83} (2011) 245117.

\bibitem{Has08}
E.~Hassinger, G.~Knebel, K.~Izawa, P.~Lejay, B.~Salce and J.~Flouquet: Phys.
  Rev. B {\bf 77} (2008) 115117.

\bibitem{Slo09}
E.~Slooten, T.~Naka, A.~Gasparini, Y.~K. Huang and A.~de~Visser: Phys. Rev.
  Lett. {\bf 103} (2009) 097003.

\bibitem{Has10}
E.~Hassinger, D.~Aoki, G.~Knebel and J.~Flouquet: J. Phys.: Conf. Ser. {\bf
  200} (2010) 012055.

\bibitem{Fay80}
D.~Fay and J.~Appel: Phys. Rev. B {\bf 22} (1980) 3173.

\bibitem{Min08}
V.~P. Mineev: J. Phys. Soc. Jpn. {\bf 77} (2008) 103702.

\bibitem{Aok09_UCoGe}
D.~Aoki, T.~D. Matsuda, V.~Taufour, E.~Hassinger, G.~Knebel and J.~Flouquet: J.
  Phys. Soc. Jpn. {\bf 78} (2009) 113709.

\bibitem{Oht10}
T.~Ohta, T.~Hattori, K.~Ishida, Y.~Nakai, E.~Osaki, K.~Deguchi, N.~K. Sato and
  I.~Satoh: J. Phys. Soc. Jpn. {\bf 79} (2010) 023707.

\bibitem{Iha10}
Y.~Ihara, T.~Hattori, K.~Ishida, Y.~Nakai, E.~Osaki, K.~Deguchi, N.~K. Sato and
  I.~Satoh: Phys. Rev. Lett. {\bf 105} (2010) 206403.

\bibitem{Aok07_NpPd5Al2}
D.~Aoki, Y.~Haga, T.~D. Matsuda, N.~Tateiwa, S.~Ikeda, Y.~Homma, H.~Sakai,
  Y.~Shiokawa, E.~Yamamoto, A.~Nakamura, R.~Settai and Y.~\={O}nuki: J. Phys.
  Soc. Jpn. {\bf 76} (2007) 063701.

\bibitem{Chu10}
H.~Chudo, H.~Sakai, Y.~Tokunaga, S.~Kambe, D.~Aoki, Y.~Homma, Y.~Haga, T.~D.
  Matsuda, Y.~Onuki and H.~Yasuoka: J. Phys. Soc. Jpn. {\bf 79} (2010) 053704.

\bibitem{Kam07b}
S.~Kambe, H.~Sakai, Y.~Tokunaga, T.~Fujimoto, R.~E. Walstedt, S.~Ikeda,
  D.~Aoki, Y.~Homma, Y.~Haga, Y.~Shiokawa and Y.~\={O}nuki: Phys. Rev. B {\bf
  75} (2007) 140509.

\bibitem{Miy86}
K.~Miyake, S.~Schmitt-Rink and C.~M. Varma: Phys. Rev. B {\bf 34} (1986)
  6554.

\bibitem{Sca86}
D.~J. Scalapino, E.~Loh and J.~E. Hirsch: Phys. Rev. B {\bf 34} (1986)
  8190.

\bibitem{Deg10}
K.~Deguchi, E.~Osaki, S.~Ban, N.~Tamura, Y.~Simura, T.~Sakakibara, I.~Satoh and
  N.~K. Sato: J. Phys. Soc. Jpn. {\bf 79} (2010) 083708.

\bibitem{Lay71}
A.~Layzer and D.~Fay: Int. J. Mag. {\bf 1} (1971) 135.

\bibitem{Mon01}
P.~Monthoux and G.~G. Lonzarich: Phys. Rev. B {\bf 63} (2001) 054529.

\bibitem{Mon07}
P.~Monthoux, D.~Pines and G.~G. Lonzarich: NATURE {\bf 450} (2007)
  1177.

\bibitem{Fuj04}
S.~Fujimoto: J. Phys. Soc. Jpn. {\bf 73} (2004) 2061.

\bibitem{Kir03}
T.~R. Kirkpatrick and D.~Belitz: Phys. Rev. B {\bf 67} (2003) 024515.

\bibitem{Sam02}
K.~V. Samokhin and M.~B. Walker: Phys. Rev. B {\bf 66} (2002) 024512.

\bibitem{Fom01}
I.~A. Formin: J. E. T. P. Lett. {\bf 74} (2001) 111.

\bibitem{Min02}
V.~P. Mineev: Phys. Rev. B {\bf 66} (2002) 134504.

\bibitem{Wal02}
M.~B. Walker and K.~V. Samokhin: Phys. Rev. Lett. {\bf 88} (2002) 207001.

\bibitem{Wat02}
S.~Watanabe and K.~Miyake: J. Phys. Soc. Jpn. {\bf 71} (2002) 2489.

\bibitem{San03}
K.~G. Sandeman, G.~G. Lonzarich and A.~J. Schofield: Phys. Rev. Lett. {\bf
  90} (2003) 167005.

\bibitem{Suh01}
H.~Suhl: Phys. Rev. Lett. {\bf 87} (2001) 167007.

\bibitem{Abr01}
A.~A. Abrikosov: J. Phys.: Condens. Matter {\bf 13} (2001) L943.

\bibitem{Mac01}
K.~Machida and T.~Ohmi: Phys. Rev. Lett. {\bf 86} (2001) 850.

\bibitem{Buz03}
A.~I. Buzdin and A.~S. Mel'nikov: Phys. Rev. B {\bf 67} (2003) 020503.

\bibitem{Tad_ICHE}
Y.~Tada, N.~Kawakami and S.~Fujimoto: arXiv:1008.4204.

\end{thebibliography}

\end{document}